# Parallel and Limited Data Voice Conversion Using Stochastic Variational Deep Kernel Learning

Mohamadreza Jafaryani[a], Hamid Sheikhzadeh[a*], Vahid Pourahmadi[a]

[a] *Department of Electrical Engineering, Amirkabir University of Technology (Tehran Polytechnic), Iran*

**Abstract**

Typically, voice conversion is regarded as an engineering problem with limited training data. The reliance on massive amounts of data hinders the practical applicability of deep learning approaches, which have been extensively researched in recent years. On the other hand, statistical methods are effective with limited data but have difficulties in modelling complex mapping functions. This paper proposes a voice conversion method that works with limited data and is based on stochastic variational deep kernel learning (SVDKL). At the same time, SVDKL enables the use of deep neural networks' expressive capability as well as the high flexibility of the Gaussian process as a Bayesian and non-parametric method. When the conventional kernel is combined with the deep neural network, it is possible to estimate non-smooth and more complex functions. Furthermore, the model's sparse variational Gaussian process solves the scalability problem and, unlike the exact Gaussian process, allows for the learning of a global mapping function for the entire acoustic space. One of the most important aspects of the proposed scheme is that the model parameters are trained using marginal likelihood optimization, which considers both data fitting and model complexity. Considering the complexity of the model reduces the amount of training data by increasing the resistance to overfitting. To evaluate the proposed scheme, we examined the model's performance with approximately 80 seconds of training data. The results indicated that our method obtained a higher mean opinion score, smaller spectral distortion, and better preference tests than the compared methods.

## 1. Introduction

Voice conversion (VC) is a system that modifies the utterance of one speaker (source) so that the listener believes that another specified speaker (target) has uttered it. A frame-level mapping VC executes conversion operations in two phases: training and conversion. In the training phase, the speakers' training data (named "parallel" or "non-parallel" depending on whether the sentences have the same linguistic content) is turned into a set of components utilizing an analysis system. Then, from the output of the analyzer, the required features, such as spectral and prosodic ones, are extracted. After that, the corresponding frames of the two speakers are discovered and aligned. In the final stage, the suitable conversion function is estimated using the aligned data. In the conversion phase, the source utterance is analyzed, and the required features are extracted; subsequently, using the conversion function obtained from the training phase, these features are converted into the target ones. Finally, the converted features are employed to synthesize the output utterance.

Spectral features contain a considerable amount of information about the speaker's identity; hence, the focus of most voice conversion methods is on spectral mapping. One of the earliest successful attempts to map spectral features was based on Vector Quantization (VQ) (Abe et al., 1990). To overcome the hard clustering problem of the VQ method, an approach using fuzzy VQ was developed (Shikano et al., 1991). The use of linear regression for each cluster of source acoustic space was proposed by (Valbret et al., 1992). Since neighboring frames may be from different clusters, this approach has the problem of temporal discontinuity. To tackle this issue, Gaussian Mixture Model (GMM) based approaches for soft clustering have been introduced. (Stylianou et al., 1998) proposed an approach that may be regarded as a hybrid of fuzzy clustering and linear conversion. (Kain and MacOn, 1998) introduced the Joint Density Gaussian Mixture Model (JDGMM) technique, in which the joint feature vectors of the two speakers are modeled by the GMM, and the conversion function parameters are extracted

---

directly from the joint distribution. The aforesaid method gained popularity in spectral mapping, but shortcomings, such as temporal discontinuities, over-smoothing, and over-fitting, are among the challenges this method encounters. In order to overcome the problem of frame-to-frame conversion, an approach based on maximum likelihood estimation was investigated that models the trajectory of spectral parameters (Toda et al., 2007).

Employing Partial Least Square (PLS) with the source GMM is another approach introduced to discover the mapping function between two speakers (Helander et al., 2010). The Dynamic Kernel Partial Least Square (DKPLS) technique was introduced to model nonlinearity as well as to retain the dynamics of the data (Helander et al., 2012). In DKPLS, the source features are converted to a higher dimension using the kernel transform, and the combination of consecutive frames is utilised. This approach has delivered good results in voice conversion with limited data.

From the outset, artificial neural networks have been used in voice conversion. Neural networks were utilized to map the formants and MFCC coefficients in (Narendranath et al., 1995) and (Baudoin and Stylianou, 1996). Due to the increasing popularity of Deep Neural Networks (DNN), they have been widely used in voice conversion. Deep neural networks (Desai et al., 2010) were applied to estimate the mapping function to transform low-dimensional spectral features. In (Nakashika et al., 2013), two distinct DBNs were utilized to extract hidden features from each speaker, and they were converted using a feed-forward network. In (Nakashika et al., 2014), the Recurrent Temporal Restricted Boltzmann Machine, a type of recurrent neural network, was utilized to include temporal and correlation information between neighboring frames. The usage of DBLSTM was also recommended to retain temporal information in (Sun et al., 2015) and (Ming et al., 2016). Applying the mechanism of attention to encoder-decoder neural networks is another proposed way in parallel voice conversion. In this approach, alignment and mapping are done concurrently. Using the attention mechanism, the length of the converted voice can be different from the source's, which can be considered a means to convert both spectral and prosodic information. Neural networks such as sequence-to-sequence conversion networks (SCENT) (Zhang et al., 2019) and AttS2S-VC (Tanaka et al., 2019) are in this class. Although very powerful in modelling and estimating the nonlinear conversion function, the methods that use deep neural networks require considerable data.

Parallel voice conversion is considered a machine learning application with limited training data. Non-parametric approaches do not apply numerous assumptions and biases to the model and give significant flexibility to the mapping function. In addition, they can demonstrate good resistance to overfitting (which is owing to the high degree of freedom of the model compared to the amount of data). Therefore, they can be regarded as a promising solution for spectral mapping with limited training data. The Gaussian process (GP) is a Bayesian and non-parametric approach whose usage in voice conversion is precedented (Pilkington et al., 2011) (Xu et al., 2014). The studied methods use exact inference, which results in computational complexity. To reduce the computational complexity, the mentioned approaches employ acoustic space clustering and train independent Gaussian processes for each cluster.

To build a successful voice conversion system, an effective conversion function must be discovered that modifies the source's identity-dependent features to fulfill the following conditions:
1. The converted speech should be of high quality, with no temporal discontinuities or over-smoothing.
2. A suitable conversion function should be trainable with a decent amount of data.
3. The identity of the target speaker must be included in the converted speech.

As our contribution, this paper introduces a voice conversion system based on Stochastic Variational Deep Kernel Learning (SVDKL) (Wilson et al., 2016a). We propose using SVDKL to discover a suitable conversion function to map the spectral features of two speakers when parallel, but limited data is available. SVDKL takes advantage of the combination of the deep neural network and the Sparse Variational Gaussian Process (SVGP) (Titsias, 2009) (Hensman et al., 2013) and is able to leverage the particular advantages of each. The source spectral features are first mapped to an intermediate vector by a neural network. This intermediate vector is then considered as the input of the SVGP. The weights and biases of the neural network are considered part of the kernel parameters and are optimized alongside the other SVGP parameters. Application of SVDKL is a novel approach in the context of voice conversion. It offers two key advantages that can ease some of the difficulties of existing systems and lead to a successful conversion.

- The Gaussian process is a powerful machine learning method and, due to its non-parametric nature, can model complex functions. One of the disadvantages of the Gaussian process is its scalability. SVGP is able to summarize the training data into some inducing variables that help reduce the computational complexity (Hensman et al., 2013). By solving the scalability problem of the model, a global conversion function for the entire acoustic space of the speakers can be estimated, which, while having the advantages of the Gaussian process, solves the problem of temporal discontinuity due to the use of different conversion functions for neighboring frames.
- SVDKL considers a deep neural network in combination with a conventional kernel as a covariance function. DNN has a great ability to find hidden representations in data but is prone to overfitting. On the other hand, standard kernel functions place a smoothness assumption on the function form, limiting their usage to complex and non-differentiable functions. Additionally, the conventional kernels such as Squared Exponential (SE) that apply Euclidean distance cause features with smaller values to be ignored, which negatively affects the quality of the converted speech. The employment of the deep neural network within the standard kernel, known as deep kernel learning (Wilson et al., 2016b), in addition to enabling the learning of spectral feature structures, results in an expressive covariance function and can ameliorate the cited disadvantages of conventional kernels. Since all model parameters are obtained via optimization of SVGP marginal likelihood that respects model complexity, SVDKL is less prone to overfitting than DNN.

The following is how this document is structured: Section 2 discusses stochastic variational deep kernel learning. Section 3 provides a high-level overview of the proposed voice conversion system. Section 4 describes the experiments and their outcomes. Finally, Section 5 concludes and brings the paper to a close.

## 2. Stochastic Variational Deep Kernel Learning

### 2.1. Gaussian process regression

A multidimensional Gaussian distribution is simply specified by a fixed-sized mean vector and a covariance matrix. The Gaussian process is a generalization of the Gaussian distribution defined over infinite space. With an analogy to Gaussian distribution, the Gaussian process is characterized by the mean function $m(x)$ and the covariance function $k(x, x')$ where $x$ and $x'$ belong to infinite space $\chi$. In practice, the normalization assumption is maintained, which means that we can expect a zero-mean function $m(x) = 0$ with a parametrized covariance function $k_\theta(x, x')$.

$$GP(m(x), k_\theta(x, x')) \quad (1)$$
$$x, x' \in \chi$$

The Gaussian process is equivalent to an S-dimensional Gaussian distribution for a finite set of size S specified in infinite space $\chi' \subset \chi, |\chi'| = S < \infty$. The Gaussian process can be viewed as a distribution over functions by using input variables as an index. Let $X = (x_1, x_2, \ldots, x_N)^T, x_i \in R^D$ and $y = (y_1, y_2, \ldots, y_N)^T, y_i \in R$ represent the observed inputs and outputs, respectively (In the case of high dimensional output, considering independence assumption, univariate Gaussian distribution can be defined for each dimension). consider observed outputs to be the sum of a Gaussian noise $N(0, \sigma_{obs}^2)$ and latent function variables $f_X = (f_{x_1}, f_{x_2}, \ldots, f_{x_N})^T$. In the presence of a GP prior on latent variables $f_X \sim GP(0, K)$, the joint distribution of training outputs and function values for a set of arbitrary inputs $X_* = (x_{*1}, x_{*2}, \ldots, x_{*T})$ is as follows:

$$\begin{bmatrix} y \\ f_{X_*} \end{bmatrix} \sim N\left(0, \begin{bmatrix} K_{XX} + \sigma_{obs}^2 I & K_{XX_*} \\ K_{X_*X} & K_{X_*X_*} \end{bmatrix}\right) \quad (2)$$

Here $K$ are matrices obtained by calculating the covariance function at the points $x \in X$ and $x_* \in X_*$. The predictive distribution defined over $f_{X_*}$ is achievable given that if the joint distribution of a collection of variables is Gaussian, then the conditional distribution is similarly Gaussian.

$$p(f_{X_*}|y, X) = N(\mu_*, \Sigma_{**}) \quad (3)$$
$$\mu_* = K_{X_*X}(K_{XX} + \sigma_{obs}^2 I)^{-1} y$$
$$\Sigma_{**} = K_{X_*X_*} - K_{X_*X}(K_{XX} + \sigma_{obs}^2 I)^{-1} K_{XX_*}$$

The Gaussian process is trained by maximizing the marginal likelihood function with respect to its hyper parameters (parameters of the covariance function $\theta$ and noise variance $\sigma_{obs}^2$).

$$\log p(y) = -\frac{1}{2} y^T (K_{XX} + \sigma_{obs}^2 I)^{-1} y - \frac{1}{2} \log|K_{XX} + \sigma_{obs}^2 I| - \frac{N}{2} \log 2\pi \quad (4)$$

The first term in (4) indicates how well the model fits the data, while the second reflects the model's complexity. These two terms compete so that the model does not overfit the data.

### 2.2. Sparse variational gaussian process

Assuming $N$ is the number of available samples, the Gaussian process's exact inference has $O(N^3)$ complexity, limiting its application to data sets with $N < 10000$ samples. To address this issue, GP with stochastic variational inference, also known as the sparse variational Gaussian process, is used [22]. Inducing inputs are employed in this approach, and the training data is summarized in some pseudo inputs and the matching function values. $Z = (z_1, z_2, \ldots, z_M)^T, z_i \in R^D$ and $f_Z = (f_{z_1}, f_{z_2}, \ldots, f_{z_M})^T, f_{z_i} \in R$ indicate inducing inputs, and their related outputs, respectively. The usage of SVGP reduces the computational complexity to $O(NM^2)$. It can address the complexity problem since the number of inducing inputs is considerably smaller than the training data $M \ll N$.

The goal of variational inference is to estimate the posterior $p(f_X, f_Z|y)$ using a variational distribution $q_\phi(f_X, f_Z)$ in such a way that Kullback-Leibler divergence (KL) between these two distributions is minimized. This minimization is equivalent to the maximization of the Evidence Lower BOund (ELBO) defined on marginal likelihood.

$$KL[q_\phi(f_X, f_z)|p(f_X, f_Z, y)] = \log p(y) + E_{q_\phi(f_X, f_z)}\left[\log \frac{q_\phi(f_X, f_z)}{p(f_X, f_z, y)}\right] \tag{5}$$

$$\log p(y) \geq ELBO = E_{q_\phi(f_X, f_z)}\left[\log \frac{p(f_X, f_z, y)}{q_\phi(f_X, f_z)}\right] \tag{6}$$

The prior distribution multiplied by the likelihood yields $p(f_X, f_Z, y)$ in Equation (6).

$$p(f_X, f_Z, y) = p(f_X|f_Z)p(f_Z) \prod_{i=1}^{N} p(y_i|f_{x_i}) \tag{7}$$

Following (Hensman et al., 2013), the variational distribution is defined as follows:

$$q_\phi(f_X, f_z) = p(f_X|f_Z)q(f_Z) \tag{8}$$
$$q(f_Z) = N(m, S) \tag{9}$$

Given that $q(f_z)$ and $p(f_X|f_z)$ are both Gaussian, the marginal distribution $q(f_X)$ will likewise be Gaussian.

$$q(f_X) = N(\mu, \Sigma) \tag{10}$$
$$\mu = \psi m$$
$$\Sigma = K_{XX} - \psi(K_{ZZ} - S)\psi^T$$
$$\psi = K_{XZ}K_{ZZ}^{-1}$$

The equation for ELBO may be simplified based on the definition of variational distribution and the relaxation of common terms related to $p(f_X|f_z)$ from the numerators and denominators of the fraction, as well as the fact that the marginal distribution related to $q(f_{x_i})$ relies solely on $x_i$.

$$ELBO = \sum_{i=1}^{N} E_{q(f_{x_i})}[p(y_i|f_{x_i})] - KL[q(f_z)|p(f_z)] \tag{11}$$

The first term of ELBO considers data fitting, while the second term acts as a measure of complexity, penalizing the ELBO as the divergence between q and p increases. The model's resilience to over-fitting is due to the second part of the above equation. An unbiased estimate for ELBO can be calculated using a small number of randomly selected training data samples (minibatch).

$$ELBO \approx \frac{N}{|B|} \sum_{i \in B} E_{q(f_{x_i})}[p(y_i|f_{x_i})] - KL[q(f_z)|p(f_z)] \tag{12}$$

When the likelihood is Gaussian, both ELBO terms can be calculated analytically. The optimal value of variational parameters $\phi = \{m, S, Z\}$, parameters related to the covariance function $\theta$ and noise variance $\sigma_{obs}^2$ can be discovered by optimizing (12) using stochastic gradient descent-based methods. After finding the optimal parameters, the distribution of the function value for the desired inputs can be obtained.

$$p(f_{X_*}|f_z) = \int p(f_{X_*}|f_z)q(f_z)dZ = N(\mu_{X_*}, \Sigma_{X_*}) \tag{13}$$
$$\mu_{X_*} = K_{X_*Z}K_{ZZ}^{-1}m$$
$$\Sigma_{X_*} = (K_{X_*Z}K_{ZZ}^{-1})S(K_{X_*Z}K_{ZZ}^{-1})^T + K_{X_*X_*} - K_{X_*Z}K_{ZZ}^{-1}K_{X_*Z}^T$$

*2.3. Deep kernel learning*

One of the essential components of the SVGP model (covered in 2.2) that greatly influences the model's ability is the definition of the covariance function (also called the kernel). This function holds all the assumptions about the function shape and defines a form of similarity or distance. Assume that $x_i$, $x_j$ are two inputs of the covariance function, and $y_i$, $y_j$ are the corresponding outputs. If $k$ demonstrates that the two inputs are similar, the associated output values of these two points will be close.

According to the mean term $\mu_{X_*} = K_{X_*,Z} K_{ZZ}^{-1} m$ in (10), it is obvious that the output is a linear combination of function values corresponding to the inducing inputs. The linear combination weights are dependent on the similarity (based on the covariance function) of the desired input with the inducing inputs $Z$.

Due to the importance of the covariance function, numerous types have been introduced. One of the most common of these is the Squared Exponential covariance function with Automatic Relevance Determination (SE-ARD).

$$k_{SE}(x_i, x_j) = \sigma_f^2 \exp\left(-\frac{1}{2}(x_i - x_j)^T \Upsilon^{-1}(x_i - x_j)\right) \tag{14}$$

The $\Upsilon = diag([l_1^2, l_2^2, \ldots, l_D^2])$ is termed the length scale, and each element $l_i^2$ is treated independently as a parameter. $\sigma_f^2$ determines the scaling factor, which indicates the variance of the function. This function employs the scaled Euclidean distance as a measure of similarity. The use of deep neural networks in tandem with conventional kernels, known as deep kernel learning (Wilson et al., 2016b), has been researched in machine learning. Simultaneous usage of the deep kernel with SVGP is recognized in the machine learning literature as Stochastic Variational Deep Kernel Learning (SVDKL) (Wilson et al., 2016a). In this technique, a feature space $H \in R^Q$ is introduced, and the regression function $G$ may be decomposed as $G = F \circ M$, in which $M: X \to H$ is the deep neural network, and $F: H \to Y$ is the SVGP, as illustrated in Fig.1. The $M$ is in charge of translating the input to the feature space $H = M(X)$, which is then used as input by the SVGP. This composition of functions is identical to an SVGP with the following covariance function:

$$\tilde{k}(x_i, x_j) = k\left(M(x_i), M(x_j)\right) \tag{15}$$

Where $k$ represents a typical kernel such as SE-ARD that acts on the Q-dimensional feature space. In this case, the kernel parameters $\theta = [\theta_{SE}, \theta_{DNN}]$ are comprised of SE-ARD parameters $\theta_{SE}$ and DNN parameters $\theta_{DNN}$. Using a deep kernel can boost the SVGP's ability to model complex functions by enhancing the expressive power and discovering data structures.

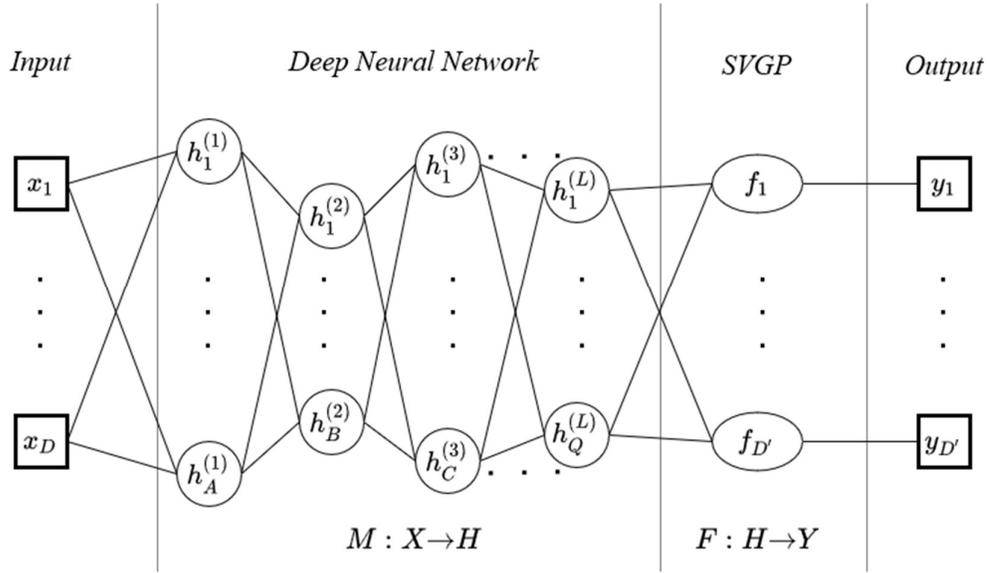

Fig. 1. Graphical representation of stochastic variational deep kernel learning

## 3. System overview

The proposed framework for voice conversion using Stochastic Variational Deep Kernel Learning is depicted in Fig.2. The proposed method uses the WORLD (Morise et al., 2016) analysis and synthesis system (a vocoder-based system) to parameterize and synthesize speech. Using WORLD, the speech signal is divided up into multiple frames, and each frame is decomposed into three different parts. 1) Fundamental frequency (F0) 2) Spectral envelope 3) Aperiodic parameters which are used to generate excitation signal. As a synthesis system, WORLD is also in charge of generating speech signals from the three previously described parts.

We employ Mel-Cepstral Coefficients (MCC) to model the Spectral Envelope. (Tokuda et al., 1994) showed that the speech spectrum could be represented using M+1 Mel-Generalized Cepstral Coefficients (MGC).

$$H(e^{j\omega}) = \begin{cases} \left(1 + \gamma \sum_{m=0}^{M} C_{\alpha,\gamma}(m) e^{-j\beta_\alpha(\omega)m}\right)^{1/\gamma}, & 0 < |\gamma| \leq 1 \\ \exp\left(\sum_{m=0}^{M} C_{\alpha,\gamma}(m) e^{-j\beta_\alpha(\omega)m}\right), & \gamma = 0 \end{cases} \quad (16)$$

Where $C_{\alpha,\gamma}$ represents spectral features. The model is determined using two parameters, $\gamma$ and $\alpha$, with $\gamma$ specifying the type of representation from all-pole ($\gamma = -1$) to cepstral ($\gamma = 0$), and $\alpha$ using the $\beta_\alpha(\omega)$ function representing the frequency warping.

$$\beta_\alpha(\omega) = \tanh^{-1} \frac{(1 - \alpha^2) \sin \omega}{(1 + \alpha^2) \cos \omega - 2\alpha} \quad (17)$$

If $\gamma = 0$ (cepstral representation) and $\alpha$ is set so that the warping equals Mel-Scale, the resulting coefficients are called MCC. Using MCC has two advantages. First, these coefficients can be simply transformed into the spectrum. The second advantage is that MCCs are relatively independent of one another, allowing us to define a separate Gaussian process for each dimension.

First, throughout the training phase, the fundamental frequency, aperiodic parameters, and spectral envelope are extracted from all frames of the speakers' utterances.
The aperiodic parameters are used directly in synthesis and are not included in training. The mean and standard deviation of the fundamental frequency in the logarithm domain are computed and saved for use in the conversion phase. Following the analysis step, 25 MCCs are extracted from the spectral envelope. The zeroth MCC represents the signal energy and is not used for training.
The dynamic time warping (DTW) algorithm is applied to the remaining 24 MCCs.
The DTW's goal is to temporally align two non-equal length sequences (source and target utterances) in order to generate aligned data sets of the source $X = (x_1, x_2, ..., x_N)^T$ and the target $Y = (y_1, y_2, ..., y_N)^T$.
Training the SVDKL is the final stage of the training phase; for each dimension of the target feature, an independent SVGP is used, and the DNN responsible for modulating the input is shared across dimensions. The model makes use of a feedforward neural network with the ReLu activation function in the middle layers and the linear activation function in the last layer. We use layerwise pretraining to accelerate the optimization process by initializing the weights of each layer in such a way that it can reconstruct the input data in its previous layer by adding a linear auxiliary layer. The Mean Square Error (MSE) is used as a network performance function in pretraining. SE-ARD is chosen as the SVGP's main kernel. The model parameters are discovered by optimizing equations (12) with the Adam optimization algorithm. The model parameters that must be optimized are as follows:

- variational parameters (per output dimension), including mean and covariance matrix of $q(f_Z)$ and inducing variables $Z$.
- covariance function parameters, including neural network parameters (weights and biases) and SE-ARD kernel parameters ($\sigma_f^2, \Upsilon$)
- Noise variance $\sigma_{obs}^2$ (per output dimension)

After extracting the MCC from the source utterance, the source features, with the exception of the zeroth coefficient, are mapped to the target ones using the optimum parameters and the mean of the predictive distribution, as given by equation (13). The converted features, along with the source's zeroth MCC, are transformed into a spectrum as stated in equation (16). Linear conversion is used in pitch mapping to transfer the mean and standard deviation of the source's F0 log-scaled to the target's one.

$$\log \hat{F}_0^y = \frac{\sigma_y}{\sigma_x} (\log F_0^x - \mu_x) + \mu_y \quad (18)$$

Where $\mu$ and $\sigma$ are the mean and standard deviation of F0 log-scaled, calculated from training data. Finally, the source aperiodic parameters, converted F0, and converted spectrum are fed into the WORLD system for synthesis, and the time domain signal is generated.

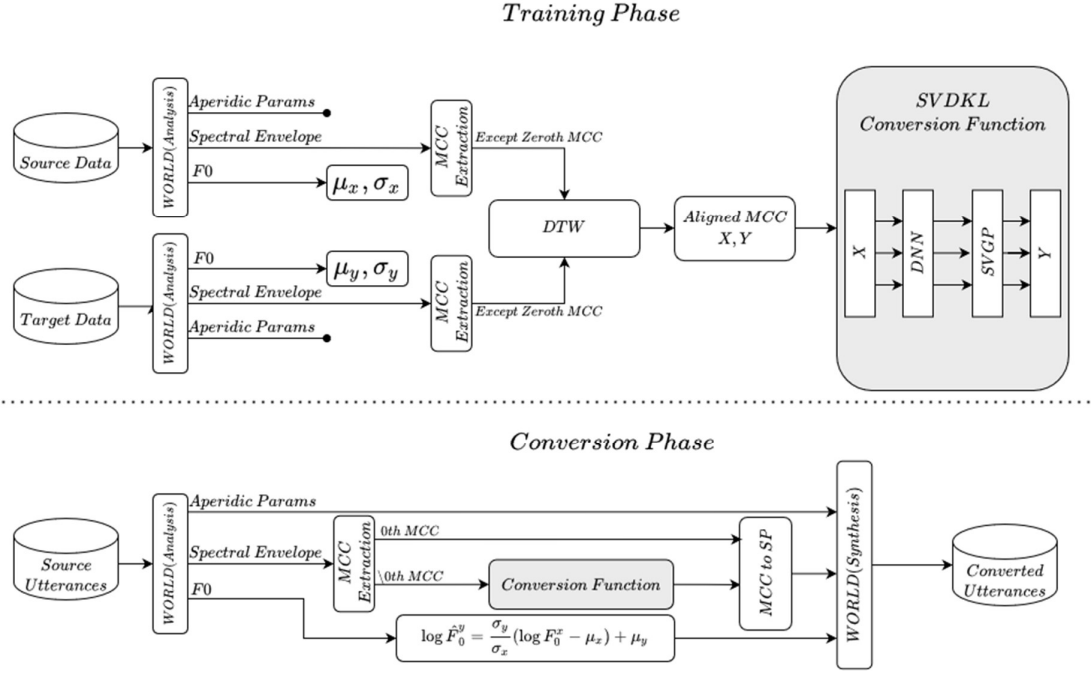

Fig. 2. Overview of Voice conversion system using stochastic variational deep kernel learning

## 4. Experimental evaluations

*4.1. Experiment setup*

The proposed method was tested using the CMU ARCTIC dataset (Kominek and Black, 2004), which contains the same utterances from different speakers. We chose four speakers: BDL and RMS, two males, and CLB and SLT, two females. For testing the ability of the proposed model in all intragender and intergender conversion pairs, four source-target pairs were considered: BDL-RMS (m-m), SLT-CLB (f-f), CLB-RMS (f-m), and BDL-SLT (m-f). Twenty sentences, equating to about 80 seconds, were chosen at random as training data. Validation data consisted of ten sentences, while test data consisted of twenty utterances. In the equation (17), the sampling frequency of 16kHz forces $\alpha = 0.41$. It's worth noting that the steps outlined in Section 3 were followed exactly for all methods used. The only distinction between them is the spectral conversion function.

*4.2. Evaluation criteria*

*4.2.1. Objective evaluation*

Mel Cepstral Distortion (MCD) is a standard measure of spectral distortion between converted and target speeches, calculated using the following equation.

$$MCD(dB) = \frac{10}{\ln 10}\sqrt{2\sum_{d=1}^{24}\left(mc_d^{(\hat{y})} - mc_d^{(y)}\right)^2} \tag{19}$$

Where $mc_d^{(y)}$ and $mc_d^{(\hat{y})}$ are the $d$th MCCs of the target and the converted speech, respectively, it should be noted that the lower MCD indicates the better performance of the voice conversion system based on this criterion.

### 4.2.2. Subjective evaluation

Subjective tests are used to gather listeners' opinions on the converted speech. In our subjective evaluation, ten different listeners (six women and four men) actively participated. For the experiments, ten sentences were chosen at random from the test sentences. In this article, the famous MOS, similarity preference, and quality preference tests were carried out.

**MOS**: Each method's sentences (all pairs) were played, and listeners were asked to rate the speech quality of each utterance on a 5-point scale (5: Excellent 4: Good 3: Fair 2: Poor 1: Bad). To reduce the listeners' bias, the order of the sentences was chosen at random from the existing methods.

**Quality preference test**: After hearing sentences converted using two different methods, test participants were asked to select the method they thought was superior in terms of quality. It is worth noting that the order of utterances is randomly chosen.

**Similarity preference test**: The natural target speech was first presented to the listeners. The outputs of the two different systems were then randomly played, and the listeners were asked to select the one that was most similar to the target.

### 4.3. Reference methods and settings

Two methods in the field of spectral mapping have been chosen to compare with the proposed method. Two factors determine the choice of two baseline methods: First, these methods are among the most successful and well-known in parallel voice conversion, and second, in the original paper, they reported system performance with limited data.

GMM-ML: a JDGMM-based method with a diagonal covariance matrix that uses Maximum Likelihood to estimate the trajectory of converted features (Toda et al., 2007).

DKPLS: a method that integrates kernel transformation, neighboring frame combining, and PLS Regression (Helander et al., 2012).

Two other methods were considered in objective evaluations to evaluate SVGP as a standalone system and compare SVDKL with the similar-architecture neural network.

SVGP: SVGP with an SE-ARD covariance function is used in the spectral mapping in this method. The number of inducing variables is the same as it is in the proposed method.

DNN: A deep neural network with the same number of layers, nodes, and activation functions as SVDKL is used in this method, but only one linear layer replaces SVGP. The early stopping is utilized to prevent the model from overfitting.

The number of Gaussians was chosen from the set of {2, 4, 8, 16, 32, 64, 128, 256} based on the minimum MCD for the validation data. In the DKPLS, the number of adjacent frames and the optimal number of latent components are also chosen to achieve the lowest MCD on the validation data. The number of inducing variables can have an impact on the quality of converted speech, and if set incorrectly, can degrade model performance. The validation MCD as a function of the number of inducing variables for the SVGP method is shown in Figure 3. Based on the results, 200 inducing variables are employed for the proposed method and SVGP.

We did early tests and discovered that at least two layers of the feed-forward neural network as the deep kernel could produce good results. In this work, we have employed a deep kernel structure similar to that provided in(Wilson et al., 2016b). The number of neurons in the last layer directly changes the lengths of inducing variables. Fig. 4 illustrates the MCD as a function of the number of neurons in the final layer, d, where the structure of the deep kernel is [1000, 500, 50, d] and the number of inducing variables is 200. Based on the observations, we chose the number of neurons in the final layer to be 20. Therefore, the final structure of the deep neural network for the kernel would be [1000, 500, 50, 20].

It's worth noting that, as shown in Figs. 3 and 4, the MCD decreases as the number of neurons and inducing variables increases, but at the expense of increasing the number of model parameters. This finding demonstrates the SVDKL model's ability to withstand overfitting. When validation data is not available, cross-validation can be used to discover the optimal number of parameters, or their number could be regarded as too large. It's important to keep in mind that adding more parameters lengthens the optimization process.

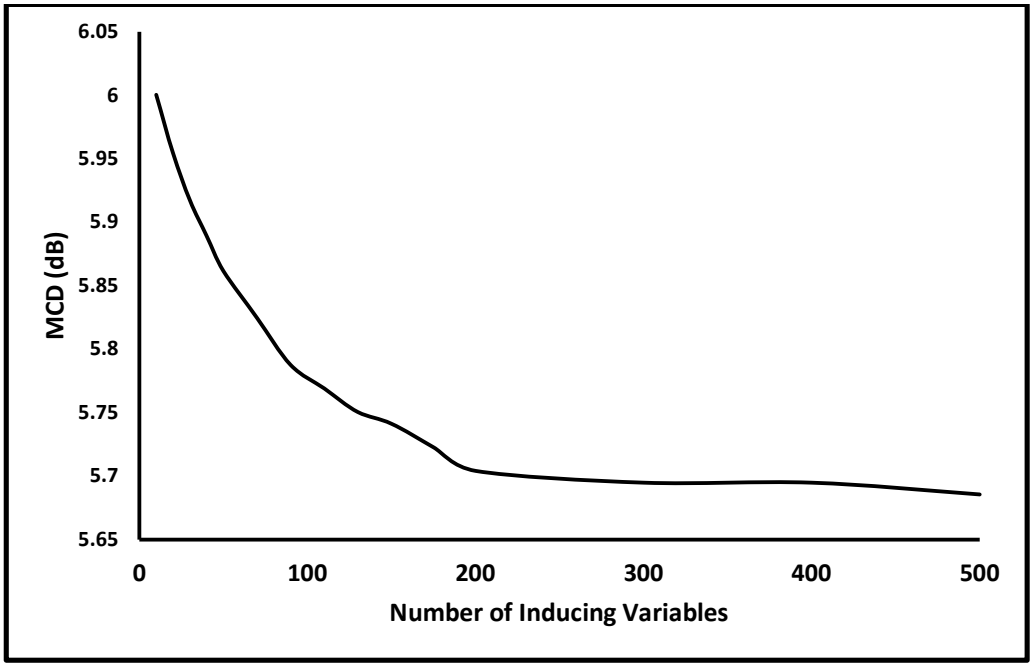

Fig. 3. SVGP average spectral distortion as a function of number of inducing variables

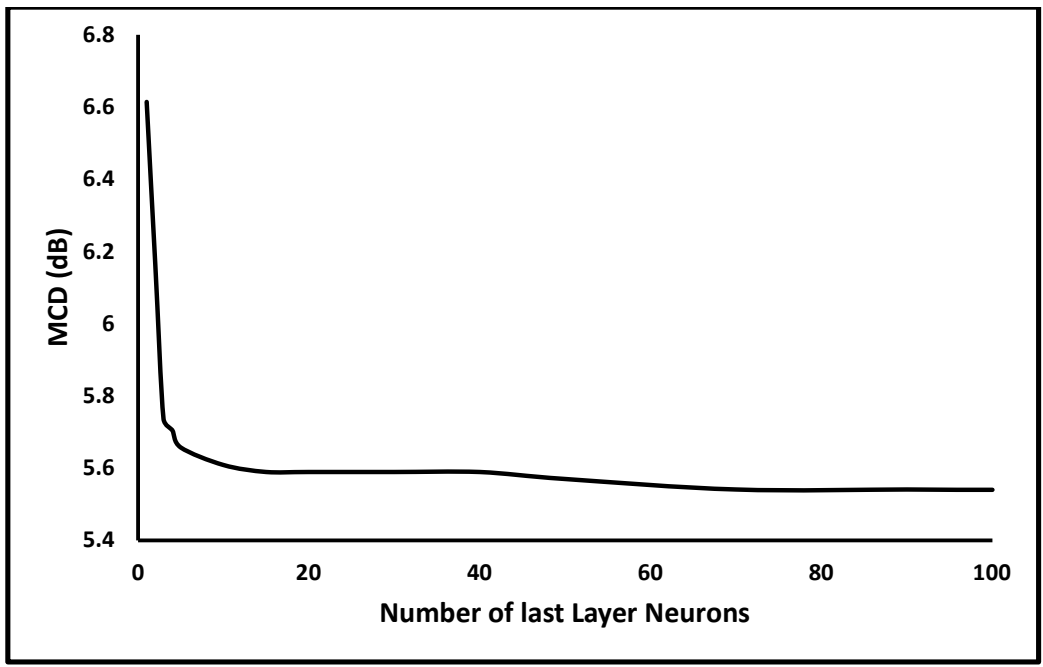

Fig. 4. SVDKL average spectral distortion as a function of number of last layer neurons when other layers follow [1000, 500, 50, d] structure and 200 inducing variables has been used

### 4.4. Results

Table 1 shows the MCD on the test data for the methods listed in Section 4.3. Given that the MCD is calculated on the dB scale, the proposed method considerably improves all conversion pairs based on objective criteria. As compared to SVGP, it can be observed that the inclusion of the deep neural network into the kernel decreases the MCD, which can be attributable to the enhanced capacity of the model to define a more accurate covariance function. The proposed method outperformed similar-structure DNN for two key reasons. First, the SVGP in the final layer of SVDKL enables the model to estimate complicated nonlinear functions with a small amount of data. Second, DNN employs MSE as a performance function for finding weights and biases. whereas SVDKL employs marginal likelihood, taking into consideration the model complexity and giving the model a stronger resilience against overfitting. The objective results show less spectral distortion between the proposed method's converted speeches than the baselines, GMM-ML, and DKPLS methods.

Table 1. Average spectral distortion for all conversion pairs using 20 training sentences. methods are described in Section 4.3

| Method | Spectral Distortion (dB) | | | | |
|---|---|---|---|---|---|
| | M→M | F→F | M→F | F→M | Average |
| SVDKL (Proposed) | **6.12** | **5.36** | **5.38** | **5.73** | **5.65** |
| DNN | 6.66 | 5.64 | 5.81 | 6.04 | 6.04 |
| SVGP | 6.17 | 5.58 | 5.48 | 5.83 | 5.77 |
| DKPLS | 6.28 | 5.53 | 5.46 | 5.84 | 5.78 |
| GMM-MLE | 6.39 | 5.49 | 5.55 | 5.90 | 5.83 |

Figure 5 shows the MOS results with the 95% confidence interval. The proposed method has an MOS score of 3.4 on average, which has been graded between fair and good according to the 5-point scale. Compared to DKPLS and GMM-ML, the proposed approach obtains a higher MOS but is still far from the natural target. It should be highlighted that analysis and synthesis are partly accountable for this quality decrease.

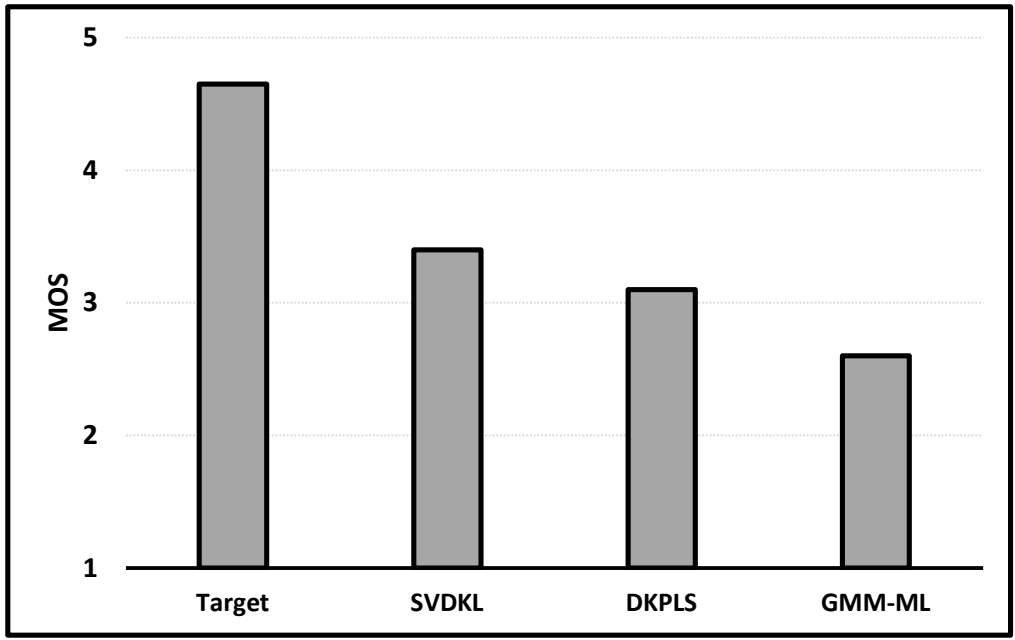

Fig. 5. Mean opinion scores with 95% confidence intervals obtained from four conversion pairs

Figure 6 demonstrates the quality preference test results between the proposed and baseline methods. On average, our approach has generated outputs of greater quality than the DKPLS in 73% of cases, and 81% of the proposed method utterances had superior quality compared to GMM-ML. The outcomes are consistent with the results obtained from MOS.

Figure 7 depicts the results of the similarity preference test. With the exception of male-to-male conversion, where there is no significant difference between our method and the DKPLS, the suggested method has produced more similar output to the target in other pairs. The output of our method was more similar to the target than the DKPLS in 69% of the cases. In intergender and intragender pairs, our method outperformed the GMM-ML method, and in 79% of the cases, it resulted in utterances more similar to the target speaker.

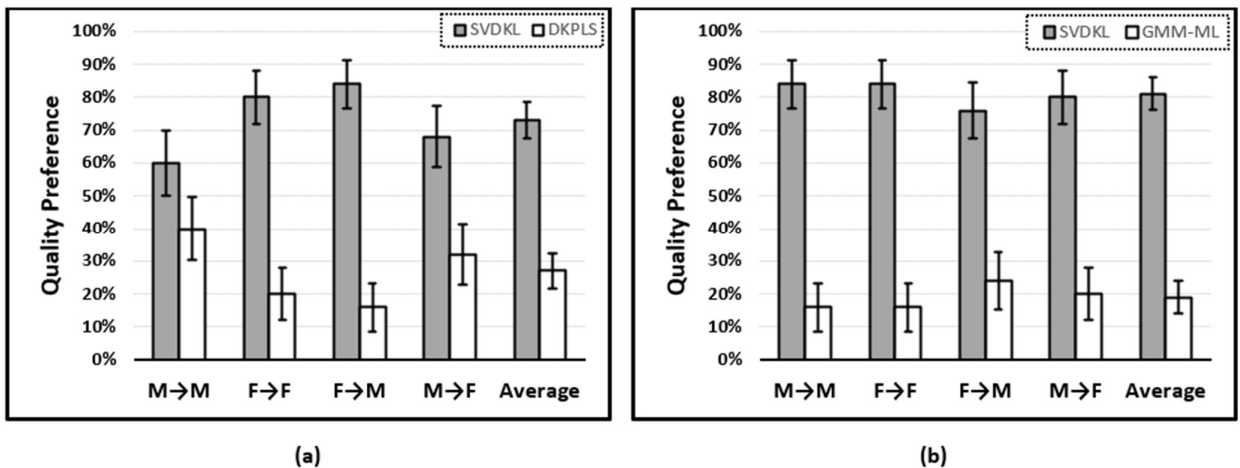

Fig. 6. The result of the quality preference test between the proposed method and (a) DKPLS, (b) GMM-ML.

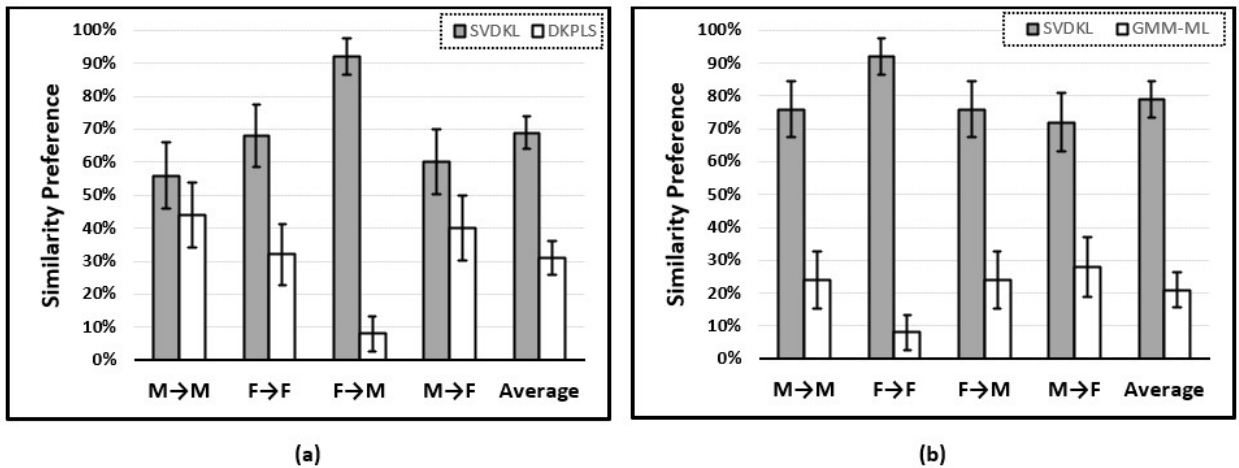

Fig. 7. The result of the similarity preference test between the proposed method and (a) DKPLS, (b) GMM-ML.

## 5. Conclusion

We proposed a voice conversion method based on stochastic variational deep kernel learning, which employs a mix of deep neural networks and sparse variational Gaussian processes. While capable of estimating complex functions, the suggested method is less prone to overfitting than deep neural networks since it learns model parameters by maximizing the marginal likelihood. A flexible and robust model eliminates the need for large-scale training data and is more in line with the spirit of voice conversion, which is a problem with limited data.

According to the results of the experiments, the proposed method produced less spectral distortion than DNN, SVGP, GMM-ML, and DKPLS. The suggested method produces better speech than the GMM-ML and DKPLS methods in terms of quality and similarity to the target. Experiments with several model configurations revealed that increasing the number of last layer neurons and the number of inducing variables reduced the MCD. The proposed method provides a novel approach to spectral mapping that makes it feasible to exploit the advantages of deep neural networks with limited data.

For future work, considering that certain papers have demonstrated quality improvements, SVDKL may be applied to convert F0 and aperiodicity in addition to spectral features. It is feasible to analyse the utilization of Neural Vocoders to boost output quality. Future research might study the employment of different kinds of neural networks in the model.

## Acknowledgements## References

Abe M, Nakamura S, Shikano K, Kuwabara H. Voice conversion through vector quantization. J Acoust Soc Japan 1990;11:71–6. https://doi.org/10.1250/ast.11.71.

Baudoin G, Stylianou Y. On the transformation of the speech spectrum for voice conversion. Int. Conf. Spok. Lang. Process. ICSLP, Proc., 1996. https://doi.org/10.1109/icslp.1996.607877.

Desai S, Black AW, Yegnanarayana B, Prahallad K. Spectral mapping using artificial neural networks for voice conversion. IEEE Trans Audio, Speech Lang Process 2010;18:954–64. https://doi.org/10.1109/TASL.2010.2047683.

Helander E, Silen H, Virtanen T, Gabbouj M. Voice conversion using dynamic kernel partial least squares regression. IEEE Trans Audio, Speech Lang Process 2012;20:806–17. https://doi.org/10.1109/TASL.2011.2165944.